\documentclass[aps,prb,preprint,superscriptaddress]{revtex4-2}
\usepackage{graphicx}  
\usepackage{dcolumn}   
\usepackage{bm}        
\usepackage{amssymb}
\usepackage[english]{babel}
\usepackage{booktabs}
\usepackage{array}
\usepackage{caption}
\usepackage{subcaption}
\usepackage{hyperref}
\hypersetup{
	colorlinks=true,
	linkcolor=green,
	filecolor=magenta,      
	urlcolor=blue,}
\begin{document}
\title{Probing pressure-induced structural evolution and its impact on the optical behaviour of vacancy-ordered halide double-Perovskite Rb$_2$TeBr$_6$}

\author{Suvashree Mukherjee}
\affiliation{Department of Physical Sciences, Indian Institute of Science Education and Research Kolkata, Mohanpur Campus, Mohanpur 741246, Nadia, West Bengal, India.}
\affiliation{National Centre for High Pressure Studies, Indian Institute of Science Education and Research Kolkata, Mohanpur Campus, Mohanpur 741246, Nadia, West Bengal, India.}

\author{Asish Kumar Mishra}
\affiliation{Department of Physical Sciences, Indian Institute of Science Education and Research Kolkata, Mohanpur Campus, Mohanpur 741246, Nadia, West Bengal, India.}
\affiliation{National Centre for High Pressure Studies, Indian Institute of Science Education and Research Kolkata, Mohanpur Campus, Mohanpur 741246, Nadia, West Bengal, India.}

\author{K.A. Irshad}
\affiliation{Elettra-Sincrotrone Trieste S.C.p.A., S.S. 14 Km 163.5 in Area Science Park,
Basovizza 34149, Italy}

\author{Bobby Joseph}
\affiliation{Elettra-Sincrotrone Trieste S.C.p.A., S.S. 14 Km 163.5 in Area Science Park,
Basovizza 34149, Italy}

\author{Goutam Dev Mukherjee}
\email [Corresponding author:]{goutamdev@iiserkol.ac.in}
\affiliation{Department of Physical Sciences, Indian Institute of Science Education and Research Kolkata, Mohanpur Campus, Mohanpur 741246, Nadia, West Bengal, India.}
\affiliation{National Centre for High Pressure Studies, Indian Institute of Science Education and Research Kolkata, Mohanpur Campus, Mohanpur 741246, Nadia, West Bengal, India.}
\date{\today}

\begin{abstract} 
The structural, vibrational, and optical properties of Rb$_2$TeBr$_6$ have been investigated under high pressure using synchrotron X-ray diffraction, Raman spectroscopy, photoluminescence (PL), and optical absorption measurements. At ambient conditions, Rb$_2$TeBr$_6$ crystallizes in the cubic Fm-3m structure, which remains stable below 8.0 GPa. {Our XRD analyses reveal that in the parent cubic phase, subtle inter-octahedral rotations develop, producing a localized deviation from the ideal cubic framework. This local asymmetry facilitates radiative recombination, leading to a pronounced enhancement of PL intensity up to 2.4 GPa. The gradual increase in phonon lifetime till about the same pressure as observed in Raman measurements corroborates an electronic transition. Above this pressure point, broadening in the widths of the Raman modes indicates an increase in the phonon scattering processes, leading to enhancement of nonradiative relaxation channels resulting in gradual PL quenching.} Additionally, the PL intensity increases upon the application of a weak external magnetic field. A structural transition to the orthorhombic Pnnm phase occurs at around 8.0 GPa, followed by a monoclinic P$2_1/m$ phase at around 10.7 GPa, and eventual amorphization at 25.5 GPa. Optical absorption spectra reveal continuous narrowing of the bandgap upon compression. These findings demonstrate the strong coupling among lattice dynamics, electronic structure, and optical response in Rb$_2$TeBr$_6$, underscoring its potential as a pressure-tunable optoelectronic material.

\end{abstract}

\maketitle

\section{INTRODUCTION}

The scientific and technological explorations of lead-free halide double perovskites have gained significant momentum in recent years, driven by the demand for stable, non-toxic alternatives to conventional lead-based optoelectronic materials \cite{{Zhang},{Ghosh}}. In this context, vacancy-ordered double halide perovskites with the general formula A$_2$BX$_6$ [where A and B are cations and X is a halide anion] have emerged as promising candidates, offering excellent stability and lead-free composition \cite{{Murugan}, {Ghorui}}. In these systems, the B site is occupied by a tetravalent cation (e.g., Te$^{4+}$), while half of the octahedral sites are systematically vacant, resulting in isolated [BX$_6$]$^{2-}$ octahedra embedded in an A-site cation framework. The isolated octahedra dominate the optical properties of perovskites, and hence are termed zero-dimensional (0D) halide perovskites \cite{Sun0D}. This distinct unit cell architecture imparts electronic, optical, and vibrational properties that differ markedly from those of corner-sharing 3D perovskite frameworks \cite{{Maughan,Ju}}.

Among this family, Rb$_2$TeBr$_6$ remains relatively unexplored but exhibits intriguing structural and optical characteristics \cite{Anazy}. It crystallizes in a cubic structure (space group Fm-3m), comprising discrete [TeBr$_6$]$^{2-}$ octahedra surrounded by Rb$^{+}$ cations, forming a vacancy-ordered double perovskite lattice \cite{{Anazy},{Abriel}}. Abriel et al. \cite{Abriel} have investigated its temperature-dependent structural evolution from 300 to 12.5 K, revealing a series of phase transitions accompanied by changes in crystal symmetry and octahedral distortions. The compound possesses a moderate indirect bandgap of ~2.1 eV, lying within the visible spectral range, and has been suggested as a potential candidate for optoelectronic applications \cite{{Anazy}, {Peresh}}. These features render Rb$_2$TeBr$_6$ a suitable model system for investigating excitonic emission and pressure-tunable optoelectronic behaviour. Despite its structural simplicity and previous insights from low-temperature studies, the optical response of Rb$_2$TeBr$_6$ under high pressure remains largely unexplored, making it an attractive material for fundamental investigation.

High-pressure techniques provide a powerful approach to probe the evolution of structural, electronic, and optical properties under lattice compression. In vacancy-ordered double halide perovskites, pressure can induce octahedral distortions, alter the band structure, and enhance radiative recombination by suppressing non-radiative pathways \cite{Mukherjee}. Previous studies on related compounds, such as Cs$_2$TeBr$_6$ and (NH$_4$)$_2$SeBr$_6$, have demonstrated pressure-induced bandgap narrowing, luminescence enhancement, and symmetry-breaking structural transitions \cite{{Samanta},{Wang}}. Beyond their fundamental interest, these findings underscore the potential of pressure as a clean and reversible means to tune optoelectronic performance without introducing chemical disorder. However, the effects of pressure on Rb$_2$TeBr$_6$ remain largely unexplored, leaving open questions about the interplay between its structural stability, bandgap evolution, and excitonic dynamics.

In this study, we systematically investigate the pressure-dependent structural and optical properties of Rb$_2$TeBr$_6$ using synchrotron X-ray diffraction (XRD), Raman spectroscopy, optical absorption, and photoluminescence (PL) measurements. Our results show that Rb$_2$TeBr$_6$ undergoes a structural transition from cubic to an orthorhombic phase close to 8.0 GPa  {followed by the monoclinic phase at about 10.7 GPa.} Under compression, the material exhibits pronounced PL enhancement, with maximum intensity at 2.4 GPa, accompanied by a gradual narrowing of the optical bandgap. {In addition, application of external weak magnetic field (0.4 T) further increases the PL intensity.} These results highlight the strong coupling between structural and electronic degrees of freedom in Rb$_2$TeBr$_6$, and reveal how lattice compression can be used to modulate its excitonic emission.

\section{Experimental Details}

Crystalline powders of Rb$_2$TeBr$_6$ are synthesized via the acid precipitation method \cite{Mukherjee}. High-purity RbBr (Purity $\geq$ 99.9\%) and TeO$_2$ (Purity $\geq$ 99.9 \%), obtained from Sigma-Aldrich, are mixed in a 2:1 molar ratio and dissolved in 2 mL of hydrochloric acid (48 wt\% in water) in a sealed glass vial. The resulting solution is stirred continuously at 110$^{\circ}$C for 2 hours, followed by an additional 2 hours of ageing at the same temperature without stirring. The mixture is then allowed to cool to room temperature and kept undisturbed overnight to facilitate crystallization. The resulting precipitate of Rb$_2$TeBr$_6$ is washed several times with ethanol and dried for further use.

Pressure-dependent Raman, photoluminescence (PL), and optical absorption measurements are performed using a piston-cylinder type diamond anvil cell (DAC) equipped with diamond anvils with 300 $\mu\text{m}$ culet \cite{Mukherjee}. A stainless-steel gasket (initial thickness 290 $\mu\text{m}$) is pre-indented to 50 $\mu\text{m}$  and a 100 $\mu\text{m}$ hole is drilled at its center to serve as the sample chamber. The Rb$_2$TeBr$_6$ sample, along with a few ruby grains for pressure calibration, is loaded into the chamber. High purity silicon oil serves as the pressure-transmitting medium (PTM). Pressure is determined using the ruby fluorescence method \cite{ruby} for Raman, PL and optical absorption measurements, whereas, for pressure calibration in high-pressure XRD measurements, a small amount of fine silver powder is added along with the sample.

Raman and PL measurements are conducted in backscattering geometry using a Monovista confocal micro-Raman spectrometer (SI GmbH) equipped with a 750 mm monochromator and a PIXIS 100BR CCD detector (1340 × 100 pixels). A 532 nm Cobolt-samba diode-pumped solid-state laser is used for excitation. The Raman measurements are performed using a Bragg filter to collect spectral data in the range 10 cm$^{-1}$ to 350 cm$^{-1}$, and the spectra are collected with a 1500 grooves/mm grating, whereas for PL measurements, 600 grooves/mm grating is used. A $20X$ long-working-distance objective is used for both focusing the laser and collecting the scattered light.

Pressure-dependent photoluminescence (PL) measurements under an external magnetic field are carried out using a miniature copper–beryllium opposing-plate diamond anvil cell (DAC) equipped with 600 $\mu\text{m}$ culet diamonds. The DAC is capable of generating pressures up to approximately 8.0 GPa. A double-pole electromagnet is employed to apply the magnetic field, providing a maximum field strength of 0.4 Tesla at the centre of the pole gap. The magnetic poles are positioned 1.5 cm apart beneath the microscope objective of the Raman spectrometer to ensure a uniform field at the sample position. This configuration allowed simultaneous high-pressure and magneto-PL measurements with excellent spectral stability and fidelity \cite{Bidisha}.

High-pressure absorption spectra are obtained using a custom-built Sciencetech 15189RD spectrometer \cite{Mukherjee}. Broadband white light is focused onto the sample using an achromatic lens (focal length 19 mm), and transmitted light is collected using a 10× objective. Absorbance (A) is calculated as:  $A= -\log(\frac { I_t -I_d}{I_0- I_d})$, where $I_0$ is the intensity of input light, $I_t$ is the intensity of light transmitted through the sample, and $I_d$ is the intensity at the dark environment. The indirect optical bandgap is determined using a Tauc plot, with the absorption edge fitted using:$(Ah\nu)^{0.5} = C \times (h\nu - E_g)$; where A is the absorbance, $h\nu$ is the energy of the photon, C is a constant, and $E_g$ is the indirect optical bandgap.

High-pressure X-ray diffraction (XRD) measurements are performed at the XPRESS beamline of the ELETTRA synchrotron facility in Trieste, Italy. A monochromatic X-ray beam with a wavelength of 0.4957 Å is focused onto the sample, utilizing a collimated beam diameter of approximately 50 $\mu\text{m}$. Diffraction patterns are collected using a PILATUS 3S 6M detector. The sample-to-detector distance is calibrated by measuring the XRD pattern of a standard LaB$_6$ sample. The 2D diffraction images are processed using DIOPTAS software \cite{dioptas}, and structural analysis is performed using GSAS with EoS fitting carried out using EOSFit and visualization using VESTA \cite{GSAS, EOSfit, Vesta}.
\section{Results}
The Rb$_2$TeBr$_6$ powder is successfully synthesized in pure phase as confirmed by the XRD measurements. The ambient XRD is indexed to a cubic structure using Fm-3m space group with lattice parameter a=10.77440(14) \AA, which is in good agreement with the literature \cite{Abriel}. The Rietveld refinement of  Rb$_2$TeBr$_6$ at ambient conditions shows an excellent fit with Rp = 1.92$\%$ and Rwp = 3.74$\%$, as shown in Figure 1(a). The refined fractional atomic coordinates are tabulated in Table I.
In the unit cell, Rb and Te atoms occupy the high-symmetry Wyckoff sites 8c (0.25, 0.25, 0.25) and 4a (0, 0, 0), respectively, whereas Br atoms are located at the 24e site with fractional coordinates (x, 0, 0), where x is a variable parameter (at ambient conditions, x=0.25028). Each Te atom is octahedrally coordinated by six Br atoms, forming TeBr$_6$ units, while the Rb atoms reside in the interstitial voids between the octahedra, exhibiting 12-fold coordination with Br under ambient conditions [Figure 1(c)].
\subsection{Pressure-dependent Photoluminescence Behavior}
At ambient pressure, Rb$_2$TeBr$_6$ exhibits a broad photoluminescence (PL) band centred at 696 nm with a full width at half maximum (FWHM) of ~160 nm under 532 nm excitation [Figure S1 in the Supplementary Information]. At ambient pressure, the PL of Rb$_2$TeBr$_6$ arises from the radiative transition between the triplet excited state ($^3P_1$) and the ground state ($^1S_0$) via self-trapped exciton (STE) states \cite{Rb2TeCl6}. Upon compression, the PL intensity enhances and reaches its maximum at 2.4 GPa, beyond which a monotonic decrease is observed [Figures 2(a) and 2(b)].
To quantitatively evaluate the pressure-induced changes in the emission profile, the PL spectra are fitted with a Gaussian function. This fitting allows us to extract the precise variation of the PL intensity, FWHM and peak position as a function of pressure. The variation of PL peak position and FWHM under pressure are shown in the Figure S2(a) and S2(b) in the Supplementary Information. At 2.4 GPa the PL intensity is nearly 120 times greater than that at ambient conditions [Figure 3]. 
In vacancy-ordered double halide perovskites, photoluminescence (PL) enhancement is frequently linked to distortions of the octahedral framework \cite{Mukherjee}. Therefore, performing pressure-dependent X-ray diffraction (XRD) measurements is crucial to elucidate the structural origin of the PL enhancement in these materials.
 
Additionally, photoluminescence measurements under external magnetic fields at various pressures are performed, which reveal a consistent enhancement in PL intensity compared with measurements performed without an applied magnetic field, while the emission peak position remains unchanged [Figure 2(c)and(d)]. {We have also performed magnetic-field-dependent PL measurements at 1.9 GPa, close to the pressure at which the PL intensity reaches its maximum, and observe a monotonic increase in intensity with increasing magnetic field [Figure S3(a) and (b) in the Supplementary Information].} The absence of any spectral shift indicates that the magnetic field does not significantly alter the electronic band structure or the nature of the emitting states. {The enhancement of photoluminescence under an applied magnetic field can be attributed to magnetic-field-induced mixing between dark and bright excitonic states \cite{Yu, Butler}. In particular, the optically inactive $^3P_0$ state, which otherwise acts as a nonradiative reservoir, acquires partial radiative character through coupling with the emissive $^3P_1$ state \cite{Taichenachev}. This leads to redistribution of the exciton population from nonradiative to radiative channels, leading to the observed increase in PL intensity.}

\subsection{Pressure-Dependent XRD}
We have performed high-pressure XRD measurements to track the evolution of the lattice and octahedral distortions, providing direct insight into the structural factors governing the observed PL behaviour in Rb$_2$TeBr$_6$. The XRD spectra of Rb$_2$TeBr$_6$ at some selected pressure points are shown in Figure S4 in the Supplementary Information. The Rietveld refinements of the XRD patterns up to 6.8 GPa indicate that Rb$_2$TeBr$_6$ preserves the cubic Fm-3m structure throughout this pressure range. Above 6.8 GPa, the X-ray diffraction pattern undergoes significant changes associated with structural instability. In particular, at 8.0 GPa the (200) peak broadens, and the (400) reflection splits, preventing reliable Rietveld refinement with the cubic symmetry. To obtain a reliable indexing of the new structure, we first analysed the XRD pattern at 9.6 GPa, where the transition appears complete. Abriel and Jhringer \cite{Abriel} have reported the low-temperature behaviour of Rb$_2$TeBr$_6$ from 300 K to 12.5 K. This study reveals that below 45 K, the compound undergoes a second-order phase transition to a tetragonally distorted structure with space group I4/m \cite{Abriel}. Motivated by this behaviour, we attempt to refine the XRD pattern at 9.6 GPa using a tetragonal model; however, this approach is unsuccessful. {A fresh indexing of the XRD pattern at 9.6 GPa leads to an orthorhombic Pnnm structure with high figure of merit having lattice parameters a = 9.81004(8) $\AA$, b = 7.21858(12) $\AA$, and c = 7.02661(16) $\AA$. For structural refinement, the initial atomic positions are obtained using the ISODISTORT software \cite{isodistort}. The Rietveld refinement yields good agreement factors (Rp = 2.71 $\%$, Rwp = 4.22 $\%$), supporting the reliability of the orthorhombic model. The corresponding Rietveld refinement profile is shown in Figure 1(b), while a schematic representation of the unit cell is presented in Figure 1(d).The refined fractional atomic coordinates at 9.6 GPa are listed in Table II. The above parameters are subsequently used to index the XRD pattern at 8.0 GPa, and the corresponding Rietveld refinement is shown in Figure S5(a) in the Supplementary Information. The relatively poor refinement reflects that 8.0 GPa lies near the onset of the structural transition.} A comparable trend is observed in the related compound K$_2$TeBr$_6$, which remains cubic (Fm-3m) at 473 K, but undergoes a sequence of structural transitions from Fm-3m to P4/mnc (tetragonal) below 445 K, followed by a transition to P2$_1/n$ in the 410–334 K range \cite{Abriel2}. Notably, Abriel \cite{Abriel2} has reported that an orthorhombic phase with Pnnm symmetry can emerge within this transition sequence, consistent with our high-pressure observations for Rb$_2$TeBr$_6$.
With further compression, the Pnnm model fails to describe the diffraction data at 10.7 GPa. The patterns can instead be refined satisfactorily using a monoclinic $P2_1/m$ phase with a = 5.84790(4) \AA, b = 5.64000(29) \AA, c = 20.89770(26) \AA, and $\beta$ = 91.536(9) $^0$. {Owing to significant peak broadening at this pressure, and absence of a suitable structural model Rietveld refinement cannot be performed; therefore, the analysis is restricted to Le Bail analysis. The Le Bail profile fit to the XRD pattern at 10.7 GPa is shown in Figure S5(b) in the Supplementary.} The stability of this monoclinic phase is confirmed up to 12.1 GPa. However, at 12.8 GPa, the diffraction peaks broaden and the $P2_1/m$ model no longer produces an adequate fit, suggesting the emergence of a new structural arrangement or coexistence of multiple distorted phases. 
Above 17.0 GPa, the diffraction patterns display progressive peak broadening, suggesting an increase in structural disorder. At 25.5 GPa, the XRD pattern is characterized by a broad diffuse feature, which indicates a pressure-induced amorphization process. Such amorphization under high pressure has been reported in other halide perovskite-related systems, and is generally attributed to the collapse of the long-range octahedral connectivity under extreme compression \cite{Lu1}. Upon decompression, the compound recovers the cubic Fm-3m structure, confirming the reversible nature of the pressure-driven transitions.

In view of the above continuous structural changes with pressure, the pressure–volume data up to only 6.8 GPa in the cubic phase are subsequently fitted using the third-order Birch–Murnaghan equation of state represented by
\begin{equation}
	P=\frac{3B_{0}}{2}\left [ \left ( \frac{V_{0}}{V} \right )^{\frac{7}{3}}-\left ( \frac{V_{0}}{V} \right )^{\frac{5}{3}}\right ]\left \{ 1+\frac{3}{4}\left ( B^{'}-4\right )\left [\left ( \frac{V_{0}}{V} \right )^{\frac{2}{3}} -1 \right ] \right \}  
\end{equation}
(where V$_0$ is the volume at zero pressure, B$_0$ is the bulk modulus and B$^\prime$
 is the pressure derivative of bulk modulus), which yields B$_0$=15.25 (12) GPa and B$^\prime$=7.70(11) [Figure 4(a)]. The pressure–volume data exhibit no anomalies up to 6.8 GPa. To examine any subtle anomalies in the lattice response under pressure, the Birch-Murnaghan equation of state is linearized as described by Polian et al. \cite{Polian}, and the resulting reduced pressure $H=\frac{P}{3f_E(1+2f_E)^\frac{5}{2}}$ is plotted against the Eulerian strain $f_E= \frac{1}{2}[(\frac{V_0}{V})^\frac{2}{3}-1]$ as an inset in Figure 4(a). Interestingly, the plot reveals a distinct change in slope at approximately 2.5 GPa, indicating a subtle alteration in the lattice response at around this pressure point. Notably, this pressure coincides closely with the observed maximum in photoluminescence intensity. To elucidate the origin of this feature, we examined the local structural geometry under compression. The TeBr$_6$ octahedra remain regular throughout this pressure range, with no detectable changes in Br–Te–Br angles, confirming the absence of internal distortion. However, the geometric relationship between neighbouring octahedra evolves gradually with pressure. The effective inter-octahedral angle, which reflects the relative orientation between adjacent TeBr$_6$ units, deviates slightly from its ideal cubic value. The deviation increases progressively with pressure up to 3.5 GPa, reaching a maximum value of about 1.8°, and then remains nearly constant (~1.5°) up to 6.8 GPa [Figure 5].

\subsection{Pressure-Dependent Raman Spectroscopy}  
Pressure dependent Raman spectroscopic measurements are carried out to probe the vibrational properties of Rb$_2$TeBr$_6$ under compression and to correlate them with the lattice evolution observed in XRD data analysis. At ambient conditions, Rb$_2$TeBr$_6$ exhibits four Raman-active modes: P$_1$ (T$_{2g}$), P$_2$ (T$_{2g}$), P$_3$(E$_{g}$), P$_4$ (A$_{1g}$). {The assignments of these modes are listed in Table III.} The P$_3$(E$_{g}$) and P$_4$ (A$_{1g}$) modes arise from the asymmetric and symmetric stretching vibrations of the Te–Br bonds, respectively, while the P$_2$ (T$_{2g}$) mode corresponds to the bending vibrations of the [TeBr$_6$]$^{2-}$ octahedra. The P$_1$ (T$_{2g}$) mode, on the other hand, is attributed to the vibrations of Rb atoms within the rigid [TeBr$_6$]$^{2-}$ framework, in line with previous studies on halide double perovskites \cite{Samanta}. However, the P$_1$ (T$_{2g}$) mode is not detectable at ambient conditions and becomes prominent under pressure at around 1.6 GPa. The Raman spectra at some selected pressure points are shown in Figure 6(a). With increasing pressure, significant modifications in the Raman spectra are observed. 

 {The P$_1$ (T$_{2g}$) mode begins to split at $\sim$8.0 GPa, accompanied by a noticeable broadening of the P$_2$ (T$_{2g}$) mode, indicating the onset of a structural transition [Figure S7 in the Supplementary Information]. The splitting of the P$_1$ and P$_2$ modes becomes clearly resolved at $\sim$9.6 GPa, signifying the completion of the cubic-to-orthorhombic phase transition. Notably, P$_4$ mode develops a shoulder that persists up to 10.2 GPa. At 11.0 GPa, the Raman spectrum shows additional modes, as indicated by arrows, suggesting a further lowering of crystal symmetry. The Raman data corroborate our XRD results, which show an orthorhombic-to-monoclinic transition at about 10.7 GPa.} 

To quantitatively analyze the pressure dependence of phonon frequencies and linewidths (FWHM), the spectra at each pressure point are fitted using Lorentzian functions. The corresponding fit to the experimental spectra at ambient conditions is presented in Figure S8 in the Supplementary. 
The pressure dependence of phonon frequencies shows distinct slope changes in linear variation near 2.4 GPa. The slope change around 2.4 GPa occurs within the cubic stability range and is not accompanied by a crystallographic phase transition.  {Since the Raman modes P$_2$ (T$_{2g}$), P$_3$ (E$_{g}$), and P$_4$ (A$_{1g}$) are associated with the vibration of TeBr$_6$ octahedra, we examined the compressibility of the octahedra to understand the observed anomaly around 2.4 GPa. At low pressure (upto $\sim$0.7~GPa), the octahedral volume follows the unit-cell compression, which exhibits a bulk modulus of 15.2(1) GPa in the cubic phase [Figure 4(a) and 4(b)]. Above $\sim$2.5~GPa, however, the bulk modulus of the octahedra increases to 23.7(4) GPa [Figure S10 in the Supplementary]. Within the quasi-harmonic approximation, the pressure dependence of the Raman frequencies can be expressed as
\begin{equation}
	\frac{d\omega}{dP} = \frac{\gamma \, \omega}{B},
\end{equation}
where $\gamma$ is the mode Gr\"uneisen parameter and $B$ is the bulk modulus. Since the bulk modulus of the octahedra increases above $\sim$2.5~GPa, the value of $\frac{d\omega}{dP}$ decreases. Therefore, the reduction in the slope of the linear pressure dependence of the Raman mode frequencies near 2.4 GPa can be attributed to the decreased compressibility of the TeBr$_6$ octahedra.
In addition to the frequency shifts, the full width at half maximum (FWHM) of the Raman modes provides further insight into the lattice response under pressure \cite{Guy,Yang}. While the mode frequencies primarily probe the evolution of bond force constants, the linewidths are sensitive to phonon lifetime, anharmonicity, and local structural disorder. Interestingly, the variations of FWHM of Raman modes in the cubic phase show nonmonotonic behaviour under pressure[Figure 6(c) and Figure S9 in the Supplementary Information]. 
The P$_4$ (A$_{1g}$) mode narrows with increasing pressure up to 2.0 GPa, above which it broadens. The FWHM of P$_3$(E$_{g}$) mode decreases with pressure up to 1.5 GPa and remains nearly constant up to 8.0 GPa. The FWHM of P$_1$ (T$_{2g}$) and P$_2$ (T$_{2g}$) modes show a minimum at 3.5 GPa with a systematic broadening thereafter. The decrease of P$_1$ (T$_{2g}$), P$_2$ (T$_{2g}$), and P$_4$ (A$_{1g}$) linewidths reflects suppression of anharmonic phonon-phonon scattering and suggest an electronic transition \cite{Samanta, Bidisha, bishnupada_prb, debabrata_prb}. Interestingly, this minimum is observed at around 2.0-3.5 GPa, where the local octahedral rotation is maximum as well as maximum in PL intensity.  }

\subsection{Pressure-Dependent optical bandgap}

The optical bandgap of Rb$_2$TeBr$_6$ is investigated as a function of pressure using UV-VIS absorption spectroscopy up to 11.4 GPa. At ambient conditions, the bandgap is found to be 2.07 eV [Figure S11 in the Supplementary], which is in good agreement with previously reported values \cite{Anazy}.
Upon compression, the optical bandgap of Rb$_2$TeBr$_6$ decreases monotonically. At ambient pressure, the valence-band maximum is primarily derived from Te-5s and Br-4p orbitals, while the conduction-band minimum mainly consists of Te-5p and Br-4p orbitals \cite{Ju}. The evolution of the optical bandgap strongly correlates with the pressure-induced changes in the Te–Br bond lengths within the TeBr$_6$ octahedra in the cubic symmetry [Figure 7(a)]. In the cubic Fm-3m phase, compression results in a uniform contraction of undistorted TeBr$_6$ octahedra, leading to a symmetric reduction of all Te–Br bond lengths. This homogeneous bond shortening enhances the overlap between Te and Br orbitals, thereby narrowing the optical bandgap \cite{Li}.
At 8.0 GPa, however, the crystal undergoes a structural transition from cubic to orthorhombic and subsequently to monoclinic symmetry, leading to distortion of the TeBr$_6$ octahedra. The distortion index (D) defined as $D=\frac{1}{n}\sum_{i=1}^{n}\frac{l_i-l_a}{l_a}$, where $l_i$ is the distance from the central atom to the ith coordinating atom, and $l_a$ is the average bond length \cite{distortion} is found to be 0.00536 at 8.0 GPa. In these low-symmetry phases, the Te–Br bonds shorten in an asymmetric manner and the octahedra exhibit tilting and bending distortions as shown in the Figure 1(d). These structural modifications further increase the orbital overlap between Te and Br, resulting in a continuous reduction of the bandgap with pressure \cite{Li}. By 6.0 GPa, the bandgap reduces to ~2.00 eV, and further decreases to ~1.85 eV at 11.4 GPa. This clear correspondence between the structural evolution and the electronic band structure highlights the crucial role of Te–Br bond dynamics in governing the pressure-dependent optical properties of Rb$_2$TeBr$_6$ . The results suggest that controlled lattice distortions can be an effective means to tune the bandgap in Te-based halide double perovskites.
Due to the spectral limitations of our setup, the optical bandgap could not be reliably determined below ~1.8 eV. Nevertheless, visual colour observations [Figure 7(b)] provide strong qualitative evidence of a continuous narrowing of bandgap under pressure. At 14.2 GPa, the sample exhibits a distinct reddish colour, consistent with the absorption edge shifting into the green region of the visible spectrum. Upon further compression to 16.2 GPa, blackish regions appear within the crystal, indicating that absorption extends across most of the visible range. By 21.0 GPa, the sample becomes nearly black, confirming broad visible absorption. These results collectively indicate that Rb$_2$TeBr$_6$ evolves into a strongly absorbing, narrow-gap semiconductor before amorphization at 25.5 GPa.

\section{Discussions}	 

{In Rb$_2$TeBr$_6$, the TeBr$_6$ octahedra and the Rb-centred cavity show different volume compression behaviour under pressure as shown in the relative volume compression in Fig-4(b). Initially, the compression of TeBr$_6$ octahedra follows the total volume change of the unit cell, but from 0.7 GPa to 2.6 GPa, it exhibits plateau-type behaviour. This indicates the sample in the pressure range 0.7 GPa to 2.6 GPa accommodates pressure primarily through relative rotations of the octahedra. Between ~2.5 and 3.5 GPa, the volume compression reorders the microscopic structure through a combination of slight octahedral rotation and volume contraction. At higher pressures, octahedral volume contraction becomes dominant and eventually leads to a structural transition from the cubic Fm-3m phase to the orthorhombic Pnnm phase near 8.0 GPa.
Consistently, the Raman spectra show a concurrent narrowing of phonon modes with a minimum in FWHM between ~2.5 and 3.5 GPa. The minimum in the FWHM of Raman modes indicate a decrease in anharmonic phonon scattering, and giving rise to an electronic transition. This is manifested in enhancement of PL intensity till about 2.4 GPa. It can be seen from the analysis of XRD data that the octahedral rotation within the cubic framework appears to act as a precursor for the enhancement of PL up to 2.4 GPa. In these vacancy ordered halide double perovskite systems, the TeBr$_6$ octahedra are separated from each other by Rb atom and act as zero dimensional system. The subtle rotation of TeBr$_6$ octahedra creates a slight local structural asymmetry, resulting in slight modification of the electrostatic equilibrium inside the cage. It strengthens the STE in this zero-dimensional system, enhancing the radiative recombination and hence increasing the PL intensity. This is evidenced in the moderate value of Huang-Rhys factor (S) at the same pressure due to an optimal electron-phonon coupling strength as calculated in Section B in the Supplementary Information.} 
In vacancy-ordered double halide perovskites, the balance between radiative and nonradiative recombination channels is strongly governed by the electron–phonon coupling strength, which can be quantified through the Huang–Rhys parameter (S). Strong electron–phonon coupling promotes phonon-assisted recombination and reduces radiative efficiency, whereas overly weak coupling facilitates thermal detrapping of carriers from self-trapped exciton (STE) states, thereby diminishing emission \cite{Rb2TeCl6}. Hence, an optimal intermediate coupling strength is essential for maximizing luminescence efficiency. The optimized S factor value is estimated to be 10.9 for Rb$_2$TeBr$_6$ [Figure S12 (b) in the Supplementary Information]. {The minimum in the Raman mode FWHM above 2.5 GPa corroborates the above effect. With further increase in pressure, the octahedral rotation becomes almost constant. However, the increase in Raman mode FWHM indicates an increase in lattice anharmonicity and hence the phonon-phonon scattering process. This increases the probability of exciton-phonon scattering, hence a decrease in PL intensity.}
In addition to pressure-induced effects, the application of an external magnetic field leads to a pronounced enhancement of the PL intensity at all investigated pressures. This behaviour provides strong evidence that spin-dependent processes play a crucial role in the emission mechanism of Rb$_2$TeBr$_6$ \cite{Butler}. In vacancy-ordered halide perovskites, self-trapped excitons often populate triplet states, which are typically nonradiative or weakly emissive due to their spin-forbidden nature. The applied magnetic field can partially lift the spin degeneracy and facilitate mixing between triplet and singlet excitonic states, thereby enhancing the probability of radiative recombination. This spin-state mixing effectively increases the emissive fraction of excitons, resulting in the observed PL intensity enhancement \cite{Morad, Zhou}. The fact that this effect persists across the entire pressure range investigated suggests that the spin–orbit and spin–lattice interactions remain active and pressure-tunable, providing an additional channel for controlling the optical response in Rb$_2$TeBr$_6$.\\
The enhancement of PL intensity under an external magnetic field indicates that Rb$_2$TeBr$_6$ could potentially be utilized as a magnetic-field-responsive luminescent material, suggesting its suitability for applications as a magneto-optical switch.
By combining complementary experimental techniques, this work provides new insights into the structure–property relationships of Rb$_2$TeBr$_6$ under pressure, with particular emphasis on excitonic recombination pathways and pressure-enhanced luminescence. Our findings deepen the understanding of vacancy-ordered double perovskites and offer valuable guidance for the design of pressure-responsive, lead-free light-emitting materials. Furthermore, this study positions Rb$_2$TeBr$_6$ as a benchmark system for probing pressure-driven phenomena in halide double perovskites, bridging the gap between structural dynamics and optoelectronic functionality.
Importantly, the insights gained here also have broader technological implications. Pressure-tunable emission and bandgap engineering in halide double perovskites open pathways for their application in next-generation light-emitting diodes (LEDs), scintillators for radiation detection, and optical sensors where high stability and lead-free composition are critical. By demonstrating how external pressure modifies the luminescence and bandgap of Rb$_2$TeBr$_6$, our work highlights strategies that could be extended to chemical or strain engineering, ultimately enabling environment friendly optoelectronic devices with tailored performance.

\section{Conclusions} 
High-pressure investigations of Rb$_2$TeBr$_6$ reveal a strong interplay between lattice distortion, vibrational dynamics, and optical properties. Within the cubic Fm-3m phase, pressure induces subtle inter-octahedral rotations that deviate from the ideal cubic geometry without distorting the individual TeBr$_6$ octahedra. This local reorientation enhances radiative recombination efficiency, reflected in the pronounced PL intensity maximum near 2.4 GPa and the concurrent narrowing of Raman linewidths. {Above this pressure point, enhanced phonon-phonon scattering provides an efficient pathway for nonradiative carrier recombination. As a result, the excitation energy is dissipated into lattice vibrations rather than emitted photons, leading to gradual PL quenching. The subsequent structural transitions occur to the orthorhombic Pnnm phase at 8.0 GPa and to the monoclinic P$2_1/m$ phase at 10.7 GPa, following a progressive disorder and amorphization above 12.8 GPa.} Overall, these results establish that pressure-driven octahedral reorientation governs the optical response of Rb$_2$TeBr$_6$ and demonstrate its potential as a model system for exploring structure–property coupling in halide double perovskites. Meanwhile, optical absorption measurements demonstrate a continuous decrease of the band gap with increasing pressure, consistent with enhanced orbital overlap under compression. These findings provide important insights into the interplay between structure, electronic states, and optical response in Rb$_2$TeBr$_6$, highlighting its potential for pressure-tunable optoelectronic applications.

\section{Acknowledgments}
The authors gratefully acknowledge financial support from the Department of Science and Technology (DST), Government of India, under the Indo-Italian Executive Programme of Scientific and Technological Cooperation, which facilitated access to the XPRESS beamline at the ELETTRA Synchrotron Light Source. The financial support from the Science and Engineering Research Board (SERB), Government of India, Grant No. CRG/2021/004343 for the development of the lab based high-pressure UV-VIS-NIR Absorption Spectrometer is gratefully acknowledged. SM also acknowledges the fellowship grant provided by the CSIR, Government of India.

\begin{table}[h]
	\caption*{Table I}{Structural parameters of Rb$_2$TeBr$_6$ in the cubic Fm-3m phase at ambient conditions obtained from Rietveld refinement.}\\
	 {Lattice parameter: a = 10.77440(14) $\AA$\\}

	\begin{tabular}{c@{\hskip 0.5in} c@{\hskip 0.5in} c@{\hskip 0.5in} c@{\hskip 0.5in} c@{\hskip 0.5in} c@{\hskip 0.5in}}
		\hline\hline 
		Atom&Wyckoff position&x/a&y/b&z/c & {$U_{iso}$(\AA)}  \\ 
				
		\hline 
		Rb & 8c & 0.25000 & 0.25000  & 0.25000 & 0.025\\
		
		 Te & 4a &0 & 0 & 0 & 0.025\\

		 Br	& 24e&  0.25028(26) & 0 & 0 & 0.025 \\

		\hline\hline
	\end{tabular}
\end{table}

\begin{table}[h]
	\caption*{Table II}{Structural parameters of Rb$_2$TeBr$_6$ in the orthorhombic Pnnm phase at 9.6 GPa obtained from Rietveld refinement.}\\
	
	Lattice parameter: a = 9.81004(8) $\AA$, b=7.21858(12)$\AA$, c=7.02661(16) $\AA$\\
	
\begin{tabular}{c@{\hskip 0.5in} c@{\hskip 0.5in} c@{\hskip 0.5in} c@{\hskip 0.5in} c@{\hskip 0.5in} c@{\hskip 0.5in}}
		\hline\hline 
		Atom&Wyckoff position&x/a&y/b&z/c & {$U_{iso}$(\AA)}  \\ 
		
		\hline 
		Rb & 4g & 0.75000 & 0.50000  & 0 & 0.025\\
		
		Te & 2a &0 & 0 & 0 & 0.025\\

		Br1	& 8h&  0 & 0.75417(36) & 0.24171(48) & 0.025  \\
		
		Br2	& 4g&  0.26300(28) & 0 & 0 &0.025 \\

		\hline\hline
	\end{tabular}
\end{table}

\begin{table}[h]
	
	\caption*{Table III}  {Origin of Raman-active phonon modes in Rb$_2$TeBr$_6$.}
	\centering
	\small
	\renewcommand{\arraystretch}{1.2}
	\begin{tabular}{|c|c|c|}
		\hline
		Raman Mode & Symmetry & Origin of Vibration \cite{Samanta}  \\
		\hline
		
		P$_1$ & T$_{2g}$ & Vibrations of Rb atoms within the rigid [TeBr$_6$]$^{2-}$ framework \\
		\hline
		
		P$_2$ & T$_{2g}$ & Bending vibration of the TeBr$_6$ octahedra \\
		\hline
		
		P$_3$ & E$_g$ & Asymmetric stretching vibration of Te--Br bonds\\
		\hline
		
		P$_4$ & A$_{1g}$ & Symmetric stretching vibration of Te--Br bonds \\
		\hline
		
	\end{tabular}
	\label{tab:raman_modes}
\end{table}

\newpage

\begin{figure}[ht]
	\centering
	\includegraphics[scale = 0.6]{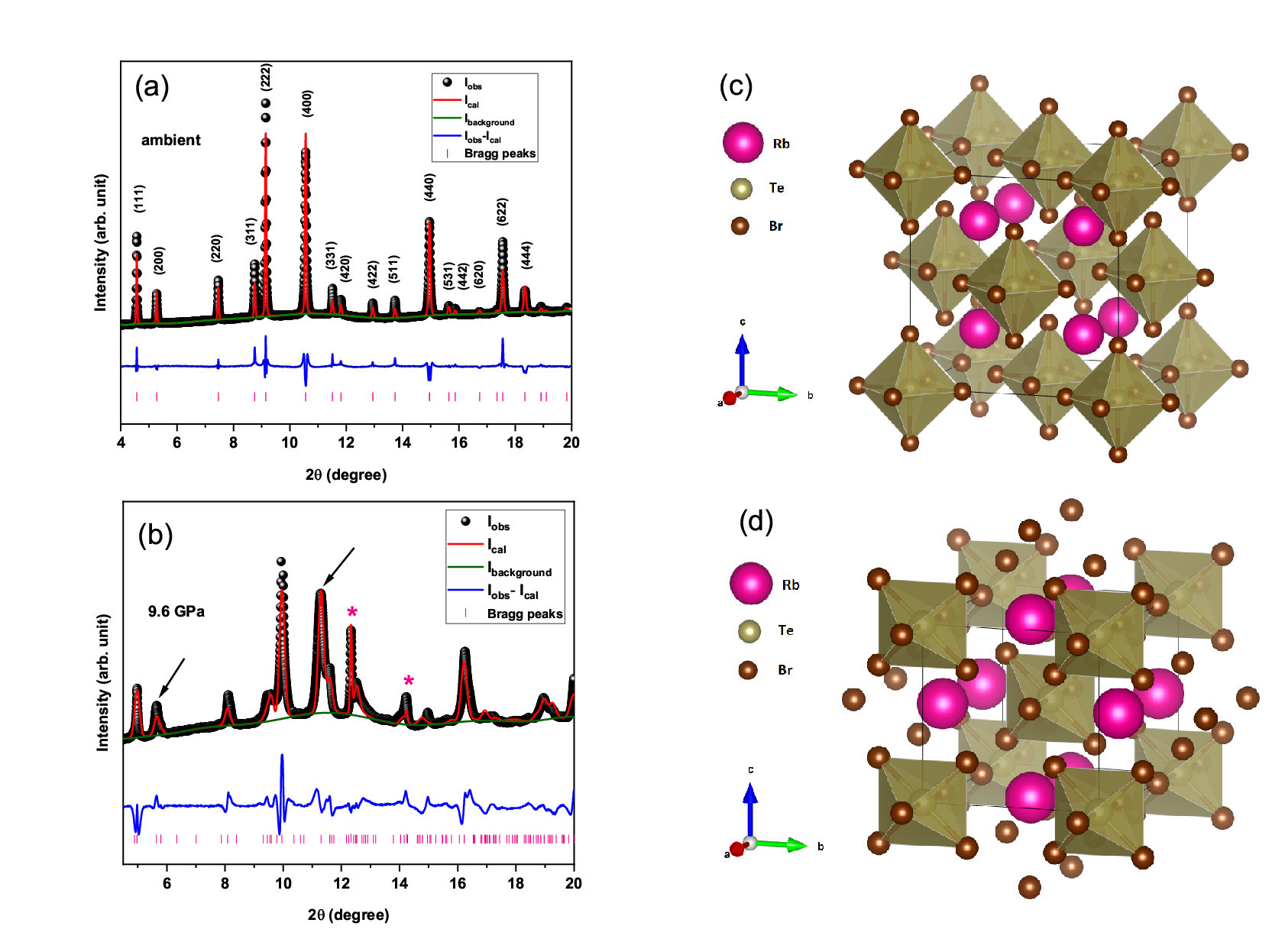}
	\caption{Rietveld refinement of the XRD pattern of Rb$_2$TeBr$_6$ at (a) ambient in cubic (Fm-3m), and (b) 9.6 GPa in orthorhombic (Pnnm) structure. The black balls represent experimental data. Red, green, and navy lines are Rietveld fit to the experimental data, background, and difference between experimental and calculated data, respectively. The magenta vertical lines show the Bragg peaks of the sample. The silver peaks at 9.6 GPa are identified with stars. The broadening of (200) peak and splitting of (400) peak at 9.6 GPa are indicated by arrows. Schematic representations of the unit cell at (c) ambient conditions, and (d) 9.6 GPa.}
	
\end{figure}
\begin{figure}[ht]
	\centering
	\includegraphics[scale = 0.6]{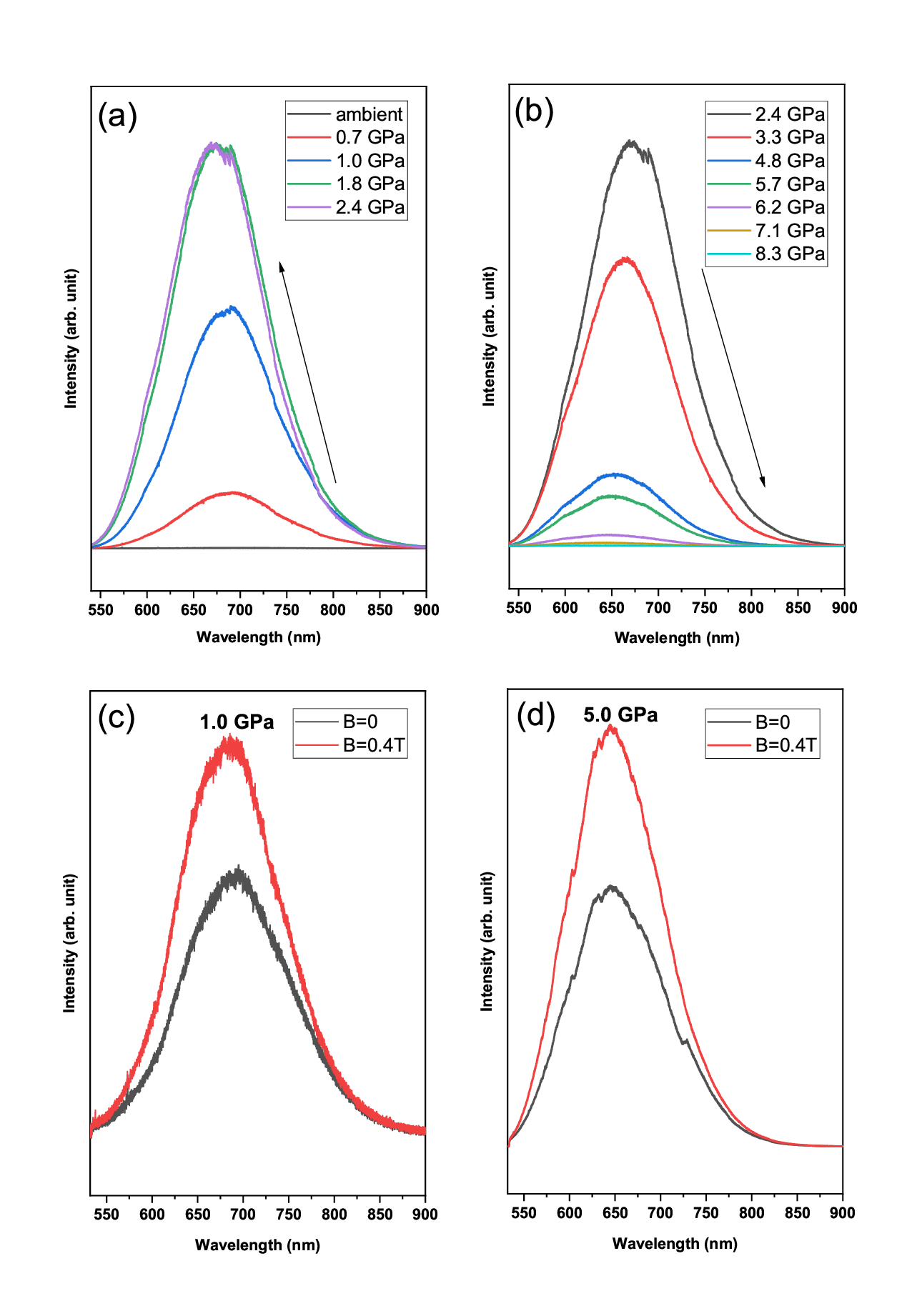}
	\caption{(a) and (b) PL spectra of Rb$_2$TeBr$_6$ at some selected pressure points.	 (c) and (d) The PL spectra of Rb$_2$TeBr$_6$ at 1.0 GPa and 5.0 GPa, respectively, in the absence and presence of an external magnetic field (0.4 Tesla). The applied magnetic flux induces a noticeable enhancement in the PL intensity.}
\end{figure}

\begin{figure}[ht]
	\centering
	\includegraphics[scale = 0.6]{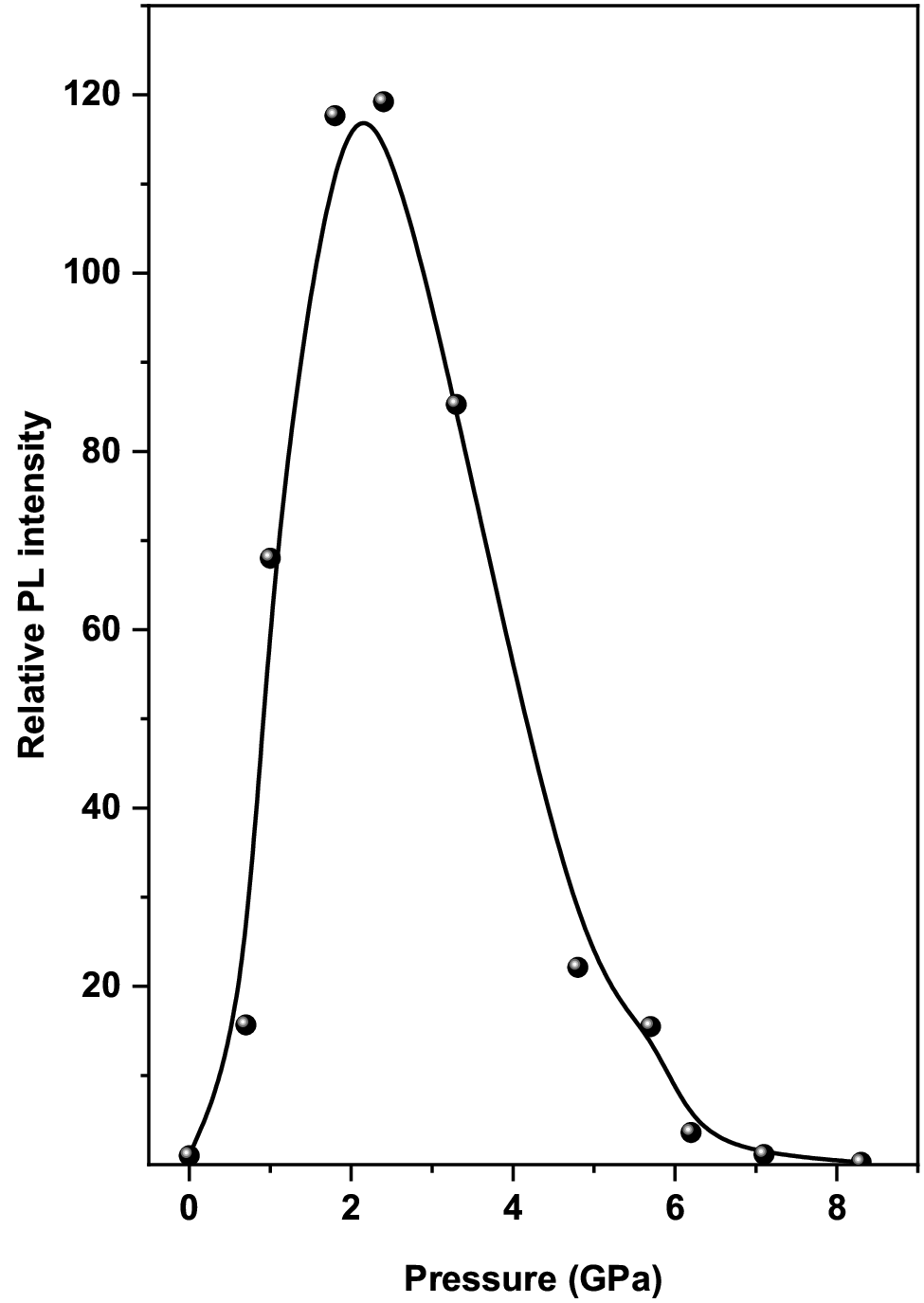}
	\caption{Relative PL intensity of Rb$_2$TeBr$_6$ under pressure. The relative PL intensity at each pressure is defined as the PL intensity at that pressure divided by the PL intensity at ambient conditions.}
\end{figure}

\begin{figure}[ht]
	\centering
	\includegraphics[scale = 0.6]{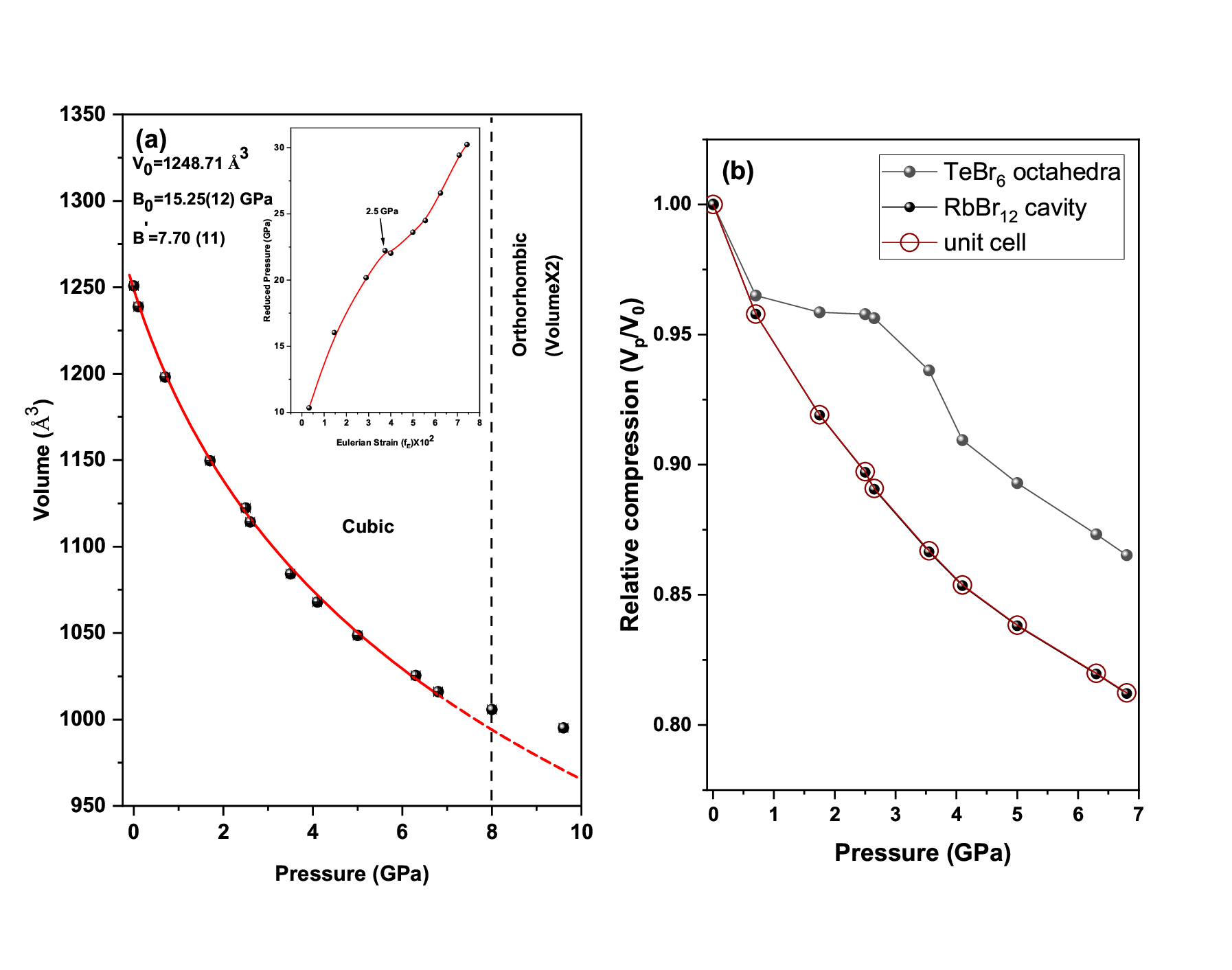}
	\caption{(a)Evolution of the unit cell volume of Rb$_2$TeBr$_6$ as a function of applied pressure in the cubic and orthorhombic phase. The red line represents the fit to the third-order Birch–Murnaghan equation of state. The inset shows a plot of  normalized pressure (H) vs Eulerian strain (f$_E$).(b)The relative volume compression of the unit cell, RbBr$_{12}$ cavity, and TeBr$_6$ octahedron under pressure in the cubic phase of Rb$_2$TeBr$_6$. The relative volume compression at each pressure is defined as the volume at that pressure (V$_p$) divided by the volume at ambient conditions (V$_0$)}
	
\end{figure}

\begin{figure}[ht]
	\centering
	\includegraphics[scale = 0.6]{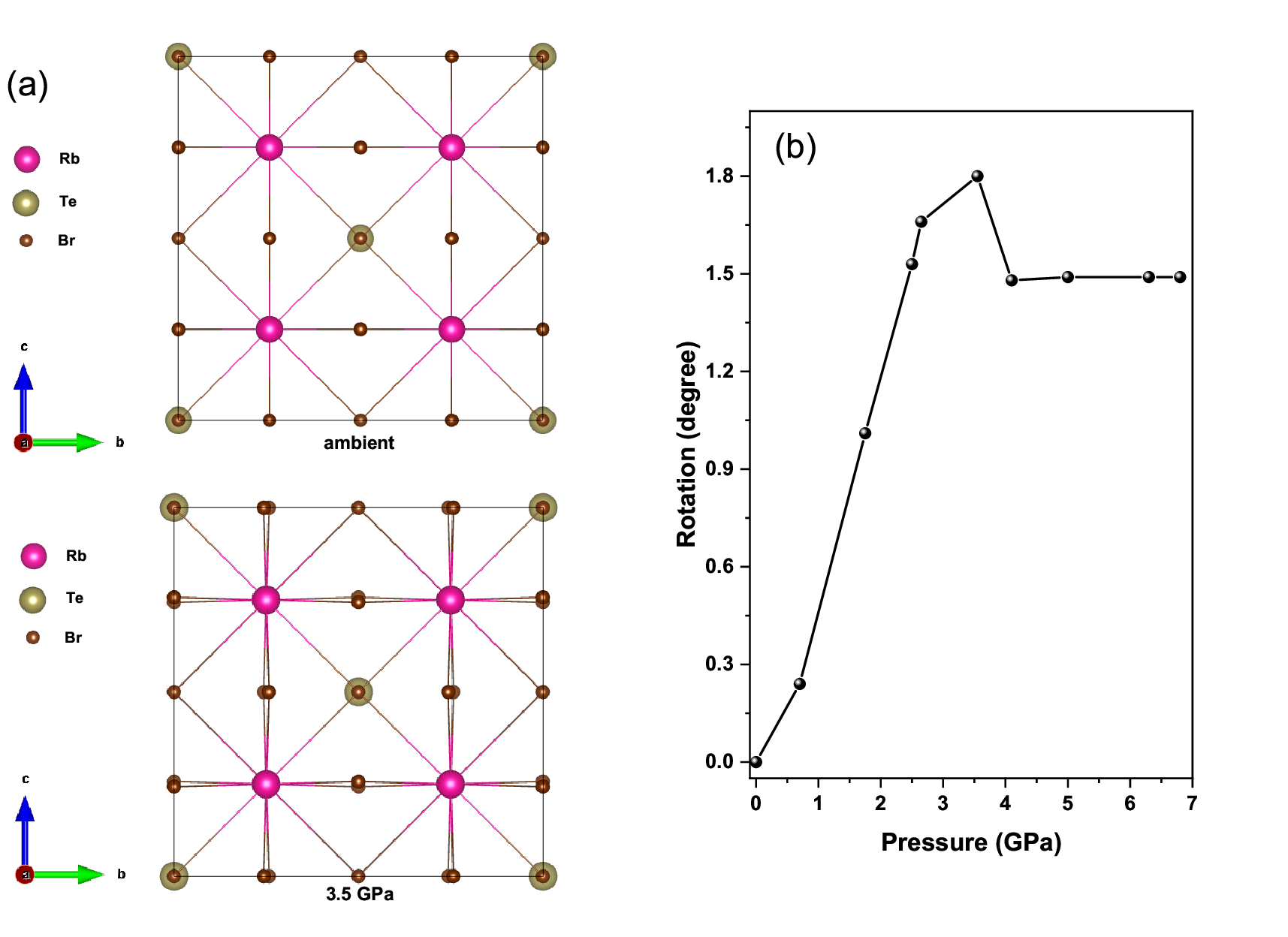}
	\caption{(a) Schematic representation of the Rb$_2$TeBr$_6$ structure at ambient pressure and 3.5 GPa, illustrating the emergence of slight octahedral rotation under compression within the cubic Fm-3m framework. (b) Pressure-induced octahedral rotational deviation in Rb$_2$TeBr$_6$ within the cubic Fm-3m phase.}
	
\end{figure}

\begin{figure}[ht]
	\centering
	\includegraphics[scale = 0.45]{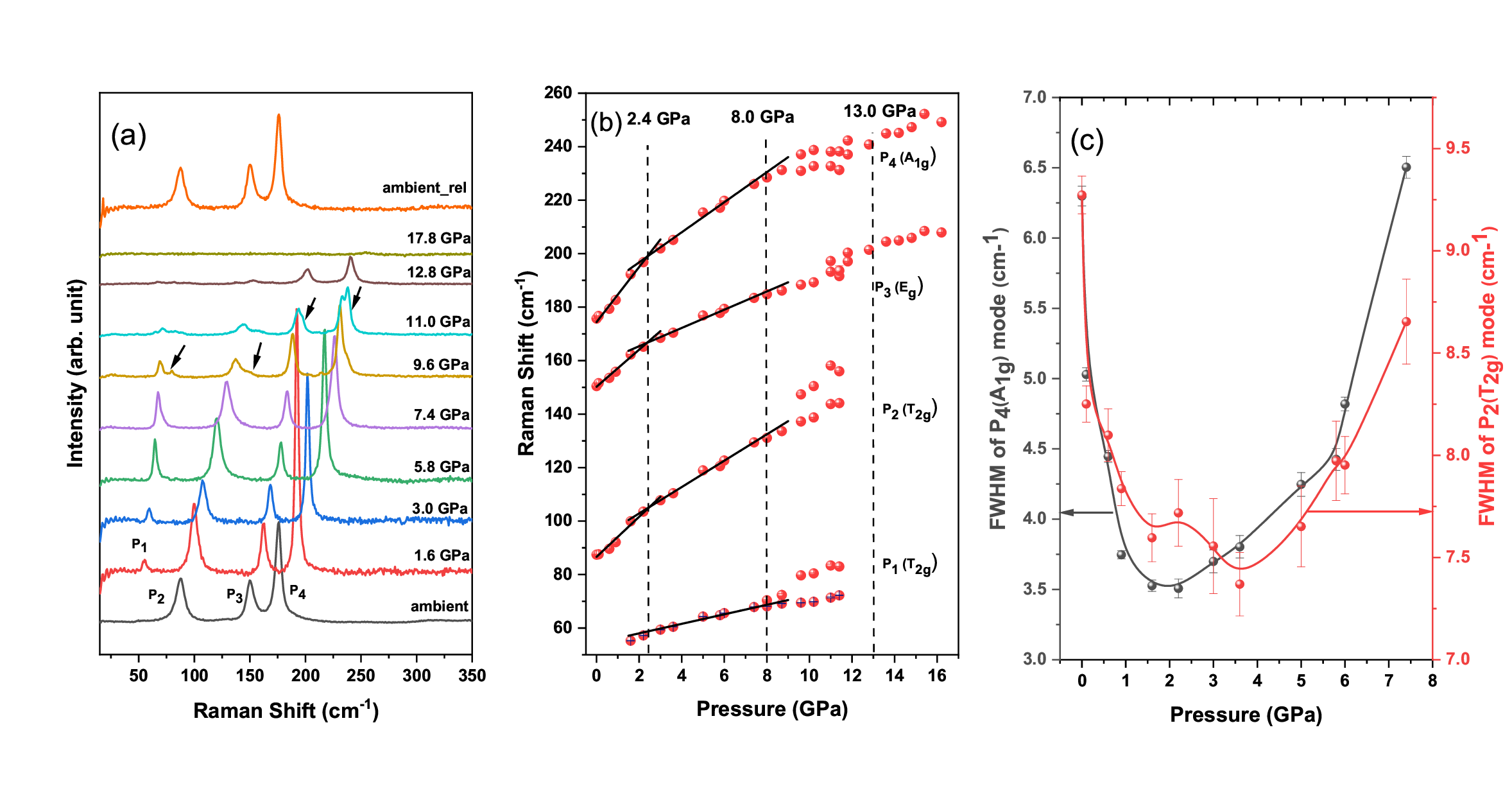}
	\caption{(a) Raman spectra of Rb$_2$TeBr$_6$ at selected pressure points. (b) Pressure evolution of Raman shift. Lines passing through the data points are the linear fit to the data.(c) Variation of FWHM of Raman modes P$_4$(A$_{1g}$) and P$_2($T$_{2g}$) with pressure at cubic phase.}
		
\end{figure}

\begin{figure}[ht]
	\centering
	\includegraphics[scale = 0.6]{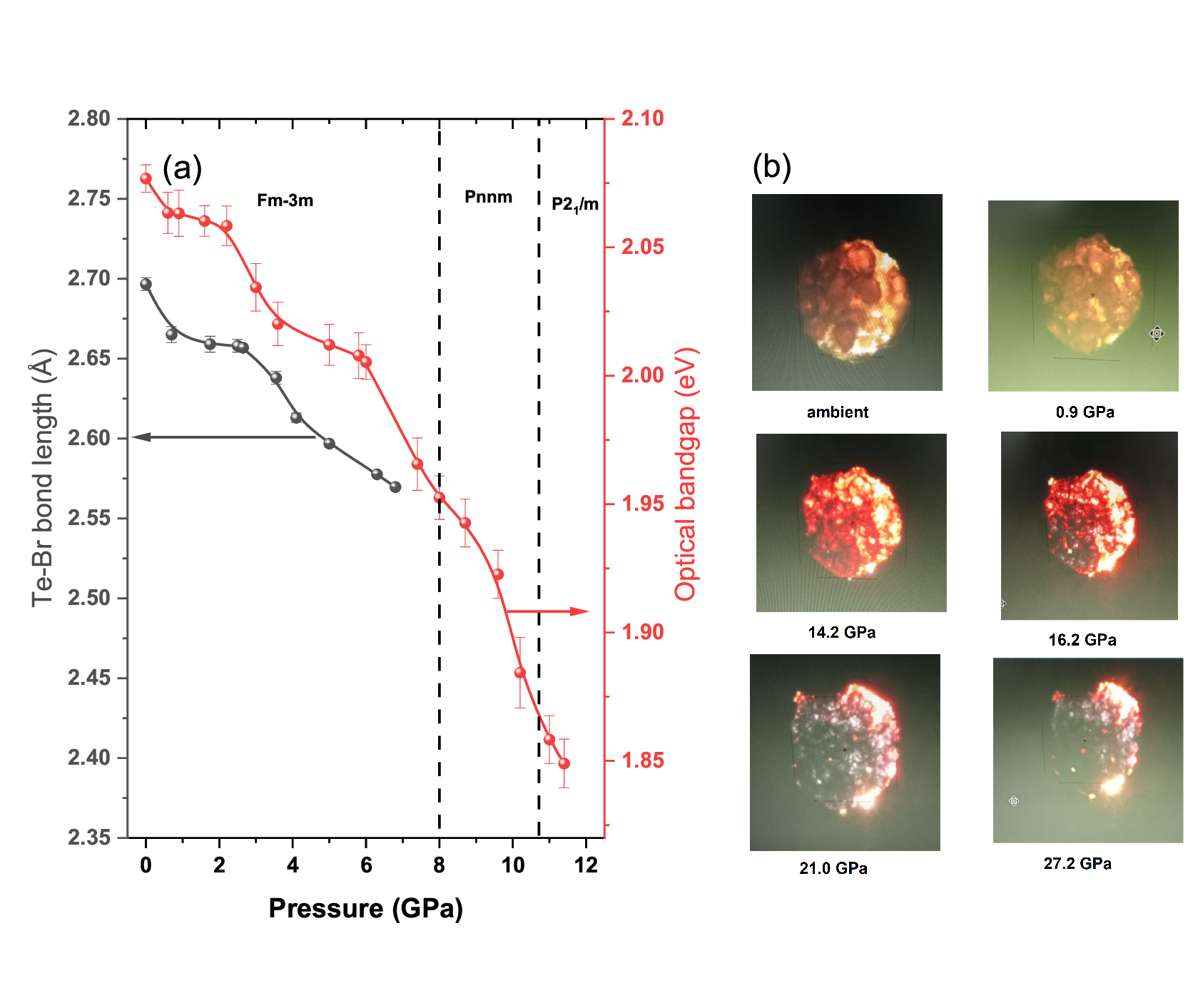}
	\caption{(a) Evolution of the optical bandgap and Te–Br bond length of Rb$_2$TeBr$_6$ as a function of pressure. The optical bandgap is shown up to 11.4 GPa, while the Te–Br bond length is presented for the cubic Fm–3m phase below 8 GPa. The dual-axis representation highlights the strong correlation between structural contraction and bandgap narrowing prior to the cubic-to-orthorhombic phase transition. (b) Optical images of Rb$_2$TeBr$_6$ under compression are shown at the right side, showing colour evolution from yellow at ambient pressure to red at 14.2 GPa and blackish at 21.0 GPa.}
\end{figure}

\end{document}